\documentclass[a4paper,fleqn]{cas-sc}
\usepackage[numbers]{natbib}
\usepackage{geometry}
\usepackage{longtable}
\usepackage{booktabs}
\usepackage{hyperref}
\usepackage{graphicx}
\usepackage{caption}
\usepackage{subcaption}
\usepackage[utf8]{inputenc}
\usepackage{ltablex} 
\def\tsc#1{\csdef{#1}{\textsc{\lowercase{#1}}\xspace}}
\tsc{WGM}
\tsc{QE}
\tsc{EP}
\tsc{PMS}
\tsc{BEC}
\tsc{DE}

\begin{document}
\let\WriteBookmarks\relax
\def\floatpagepagefraction{1}
\def\textpagefraction{.001}
\shorttitle{Predicting poverty via boosting algorithms}
\shortauthors{E.L.V. Salvador}

\title [mode = title]{Use of Boosting Algorithms in Household-Level Poverty Measurement: A Machine Learning Approach to Predict and Classify Household Wealth Quintiles in the Philippines
}                      

\author[1,2]{Erika Lynet V. Salvador}[type=editor,
                        auid=000,bioid=1,
                        prefix=Ms.,
                        role=Researcher,
                        orcid=0009-0003-2045-5459]
\cormark[1]
\fnmark[1]
\ead{erisalvador28@amherst.edu}
\ead[url]{linkedin.com/in/salvadorerika/}

\credit{Conceptualization, Data Curation, Formal Analysis, Investigation, Methodology, Software, Visualization, Writing – Original Draft, Writing – Review & Editing}

\affiliation[1]{organization={Department of Mathematics and Statistics, Amherst College},
                addressline={College St.}, 
                city={Amherst},
                postcode={01002}, 
                state={Massachusetts},
                country={United States of America}}

\affiliation[2]{organization={Science, Technology, Engineering, and Mathematics Strand, Senior High School Department, De La Salle University Integrated School Manila},
                addressline={Taft Avenue, Malate}, 
                city={Manila},
                postcode={2401}, 
                state={National Capital Region},
                country={Philippines}}

\cortext[cor1]{Corresponding author}

\begin{abstract}
This study assessed the effectiveness of machine learning models in predicting poverty levels in the Philippines using five boosting algorithms: Adaptive Boosting (AdaBoost), Cat Boosting (CatBoost), Gradient Boosting Machine (GBM), Light Gradient Boosting Machine (LightGBM), and Extreme Gradient Boosting (XGBoost). CatBoost emerged as the superior model and achieved the highest scores across accuracy, precision, recall, and F1-score at 91\%, while XGBoost and GBM followed closely with 89\% and 88\% respectively. Additionally, the research examined the computational efficiency of these models to analyze the balance between training time, testing speed, and model size—factors crucial for real-world applications. Despite its longer training duration, CatBoost demonstrated high testing efficiency. These results indicate that machine learning can aid in poverty prediction and in the development of targeted policy interventions. Future studies should focus on incorporating a wider variety of data to enhance the predictive accuracy and policy utility of these models.
\end{abstract}

\begin{highlights}
\item CatBoost achieved the highest accuracy (90.93\%) and overall performance metrics, followed by XGBoost, GBM, and LightGBM. AdaBoost had the lowest performance.
\item AUC-ROC scores indicated that CatBoost, GBM, LightGBM, and XGBoost excelled in distinguishing between poverty classes, while AdaBoost lagged behind.
\item Computational efficiency varied, with AdaBoost having the shortest training time but longest testing time. CatBoost had the longest training time but was highly efficient during testing. GBM, LightGBM, and XGBoost balanced well between training and testing times.
\end{highlights}


\maketitle

\section{Introduction}

As of 2024, over 700 million people globally live in extreme poverty and survive on less than \$2.15 (Php 125) per day \citep{worldbank2024poverty}. To address this, governments worldwide are intensifying efforts to achieve Sustainable Development Goal 1 (SDG) which targets the eradication of poverty in all its forms by 2030. However, recent research suggests that the lingering effects of the COVID-19 pandemic may endure in various countries until 2030 \citep{shulla2021effects}. This presents a significant challenge to the goal of reducing global poverty—a target already at risk prior to the crisis. Hence, the need for significant political intervention is now more pressing than ever \citep{worldbank2024poverty}. In order for policymakers to formulate targeted interventions and allocate resources efficiently, accurately determining poverty levels is paramount \citep{riegner2016implementing}. Data empowers governments and organizations to devise strategies that genuinely uplift individuals and communities from poverty. Without accurate data, policy initiatives risk falling short in addressing the underlying causes of poverty or reaching those most in need \citep{grindle2004good}.

Poverty, however, is defined diversely. Broadly, poverty measurement approaches fall into two categories: monetary and non-monetary \citep{decerf2023preference}. The monetary approach, as the name suggests, defines poverty based on income or expenditure. For instance, the established poverty methodology in the Philippines employs pre-tax income as an indicator of household well-being \citep{albert2023analysis, briones2021income}. On the other hand, several researchers contend that poverty extends beyond financial lack and includes aspects such as opportunity, education, and healthcare deficits \citep{alkire2015multidimensional}. Simply put, they view poverty as multidimensional, not solely defined by money.

Unfortunately, the drawback of conventional econometric methodologies in forecasting poverty is their tendency to oversimplify the multidimensional nature of poverty. Many prevailing measurements for poverty are structured without consideration for non-monetary indicators of welfare, such as an individual’s and household’s health, nutritional, or educational status \citep{watson2017non}. Econometric models often rely on pre-selected features (i.e., income) based on economic theory or prior knowledge, which may often overlook important relationships or interactions in the data. Therefore, it is of vital importance to conduct assessments considering diverse dimensions of poverty to formulate effective reduction policies.

Efforts to refine poverty measurement increasingly employ diverse data and machine learning methodologies. Machine learning models present distinct advantages over econometric counterparts because they can mitigate multicollinearity, achieve heightened accuracy, process data expeditiously, accommodate extensive datasets, and minimize human involvement \citep{shobana2021forecasting}. Furthermore, machine learning algorithms excel in feature selection by automatically identifying relevant variables and capturing data relationships, even amidst nonlinearity or obscured patterns \citep{li2017feature}. Each variable’s impact on poverty is scrutinized during selection and favors those with significant effects to construct models. These capabilities allow machine learning models to effectively handle high-dimensional data and generate more accurate predictive models.

Only a limited number of studies have employed machine learning methods to address poverty in the Philippines, utilized nationwide data from the Demographic and Health Survey (DHS) Program and compared various these machine learning models. For example, one study \citep{tingzon2019mapping} utilized geospatial data, including nighttime lights and daytime satellite imagery, to model socioeconomic well-being, which achieved an R\textsuperscript{2} (goodness of fit) of 0.63. Similarly, another study \citep{ledesma2020interpretable} used social media data, low-resolution satellite images, and volunteered geographic information to map poverty, which obtained an R\textsuperscript{2} of 0.66 compared to 0.63 with satellite imagery alone. These studies primarily relied on geospatial data. Conversely, another research \citep{talingdan2019performance} analyzed a dataset from the Community-Based Monitoring System (CBMS) Database of Lagangilang, Abra, utilizing various models but with only 13 features, where the Naive Bayes model yielded the best performance. Another related study \citep{repollo2021applying} applied K-means clustering and 17 features within a small, undisclosed community in the Philippines.

Significantly, a gap exists in the utilization and comparison of various machine learning methodologies for assessing poverty in the Philippines. None of the previously mentioned studies explored extensive datasets comprising hundreds of household characteristic features suitable for model estimation. While \cite{tingzon2019mapping} utilized data from the DHS, they focused on only four main features. Therefore, this study aimed to broaden poverty analysis using machine learning techniques across an extensive dataset, working under less restrictive assumptions to effectively identify low-income households. Furthermore, this research aimed to tackle the underutilization of boosting algorithms in poverty prediction \citep{li2022poverty}. In fact, XGBoost is the sole boosting algorithm with a 14\% usage rate in a recent scoping review on machine learning and poverty prediction \citep{usmanova2022utilities}. However, recent research indicates that the family of boosting algorithms has expanded with several compelling proposals and boasts improvements in both speed and accuracy \citep{bentejac2021comparative}. As a result, this study sought to utilize a subset of boosting algorithms in predicting poverty in the Philippines.

\section{Data and Methods}

\subsection{Data Cleaning}
The data for this study was obtained from the 2022 DHS in the Philippines. The original dataset consisted of 2,099 features collected from 30,372 households. A threshold of 3,050 was assigned (Figure X) because most features had fewer than 3,050 missing values. Columns exceeding the missing value threshold were removed, and rows with any remaining null values were deleted. Furthermore, some interview-related logistic features, such as the date of the interview, were manually removed. After this step, the dataset was reduced to 396 features from 2,099 features, and 20,679 households from 30,372 households.

\begin{figure}[htbp]
    \centering
    \includegraphics[width=0.6\textwidth]{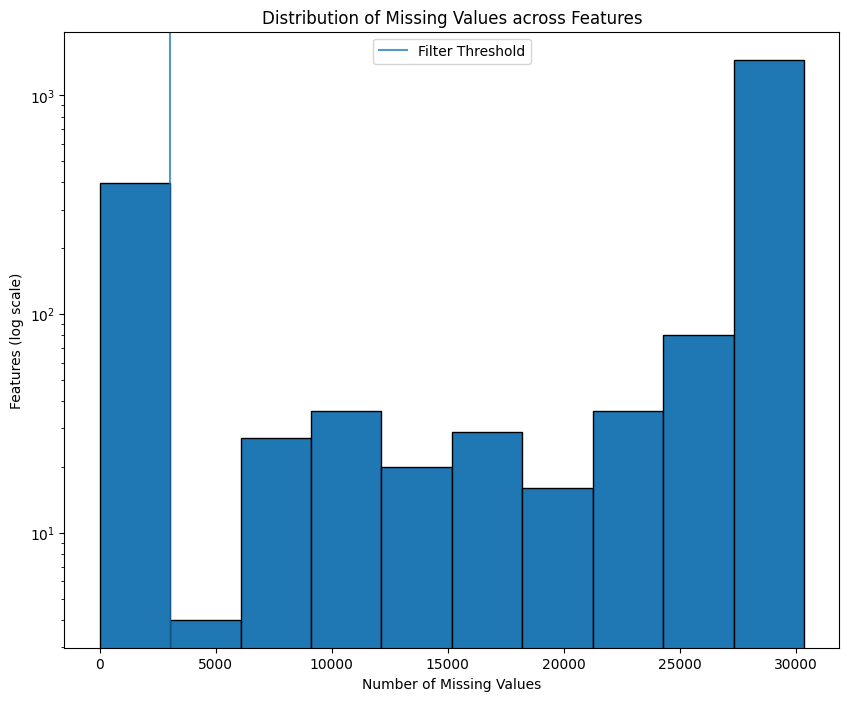} 
    \caption{Distribution of Missing Values across Features. Blue Line = 3,050}
\end{figure}

\subsection{Data Partitioning}
After data cleaning, the dataset underwent random partitioning, with 80\% allocated for training and the remaining 20\% reserved for assessing model performance. Additionally, to optimize the model's hyperparameters, a validation set comprising 10\% of the training set was required. This division was executed using a stratified sampling technique.

\subsection{Feature Scaling}
Afterwards, scaling was implemented to ensure uniform scaling across all features. Binary features were left unscaled, while numerical and ordinal features were scaled. First, the scaler was trained on the training data using z-score normalization. The training data was then transformed according to these normalization parameters. Finally, the same parameters were applied to the testing data. This ensured consistency and prevented data leakage.

\subsection{Feature Selection}
To address potential efficiency issues associated with an excessive number of features \citep{li2017feature}, the study employed the SelectFromModel() method to identify significant features for each model. SelectFromModel is a feature selection technique in scikit-learn that acts as a meta-transformer compatible with any estimator that evaluates feature importance, such as decision trees or L1-norm regularized linear models. After evaluating all models, the study tallied the frequency of each feature's selection and identified those most frequently chosen as the final set of selected features. This process resulted in the selection of 66 features deemed most relevant for predicting poverty. Pearson’s Correlation Coefficient was applied to the subset of selected features to check for multicollinearity. For pairs of features with a correlation coefficient of 0.8 or higher, the feature with the lower selection frequency was removed. From the original 36 features initially selected via SelectFromModel(), the final count remained the same, as there is minimal to no multicollinearity among them. These features, ranked by their frequency in the SelectFromModel() results, are listed in Table 1.

\begingroup
\begin{longtable}{c p{5cm} p{9cm}}
    \caption{Description of Features} \\
    \toprule
    \textbf{No.} & \textbf{Feature} & \textbf{Description} \\
    \midrule
    \endfirsthead
    \toprule
    \textbf{No.} & \textbf{Feature} & \textbf{Description} \\
    \midrule
    \endhead
    \bottomrule
    \endfoot
    \bottomrule
    \endlastfoot
    1 & Source of drinking water & Can take on 20 different values from ‘piped water’ (10), ‘tube well water’ (20), ‘dug well’ (30), ‘surface from spring’ (40), and more. \\ \midrule
    2 & Type of toilet facility & Can take on 17 values from ‘flush toilet’ (10), ‘pit toilet latrine’ (20), ‘no facility’ (30), and more. \\ \midrule
    3 & Has television & Can take on values ‘yes’ (1) or ‘no’ (0). \\ \midrule
    4 & Has refrigerator & Can take on values ‘yes’ (1) or ‘no’ (0). \\ \midrule
    5 & Has bicycle & Can take on values ‘yes’ (1) or ‘no’ (0). \\ \midrule
    6 & Has motorcycle/scooter & Can take on values ‘yes’ (1) or ‘no’ (0). \\ \midrule
    7 & Has car/truck & Can take on values ‘yes’ (1) or ‘no’ (0). \\ \midrule
    8 & Main floor material & Can take on 15 different values from ‘natural’ (10), ‘rudimentary’ (20), ‘finished’ (30), and more. \\ \midrule
    9 & Main wall material & Can take on 22 different values from ‘natural’ (10), ‘rudimentary’ (20), ‘finished’ (30), and more. \\ \midrule
    10 & Main roof material & Can take on 18 different values from ‘natural’ (10), ‘rudimentary’ (20), ‘finished’ (30), and more. \\ \midrule
    11 & Has telephone (landline) & Can take on values ‘yes’ (1) or ‘no’ (0). \\ \midrule
    12 & Type of cookstove & Can take on 11 different values from ‘electric stove’ (1), ‘solar cooker’ (2), ‘liquefied petroleum gas (LPG)/cooking gas stove’ (3), and more. \\ \midrule
    13 & Share toilet with other households & Can take on values ‘yes’ (1) or ‘no’ (0). \\ \midrule
    14 & Type of cooking fuel & Can take on 13 values from ‘alcohol/ethanol’ (1), ‘gasoline/diesel’ (2), kerosene/paraffin (3), and more. \\ \midrule
    15 & Location of toilet facility & Can take on 11 different values from ‘electric stove’ (1), ‘solar cooker’ (2), ‘liquefied petroleum gas (LPG)/cooking gas stove’ (3), and more. \\ \midrule
    16 & Has watch & Can take on values ‘yes’ (1) or ‘no’ (0). \\ \midrule
    17 & Has a computer & Can take on values ‘yes’ (1) or ‘no’ (0). \\ \midrule
    18 & Has bank account & Can take on values ‘yes’ (1) or ‘no’ (0). \\ \midrule
    19 & Type of light at home & Can take on 16 different values from ‘electricity’ (1), ‘solar lantern’ (2), ‘rechargeable flashlight’ (3), and more. \\ \midrule
    20 & Has washing machine & Can take on values ‘yes’ (1) or ‘no’ (0). \\ \midrule
    21 & Has air conditioner & Can take on values ‘yes’ (1) or ‘no’ (0). \\ \midrule
    22 & Has gas range/stove with oven & Can take on values ‘yes’ (1) or ‘no’ (0). \\ \midrule
    23 & Has microwave/toaster oven & Can take on values ‘yes’ (1) or ‘no’ (0). \\ \midrule
    24 & Has audio component/karaoke & Can take on values ‘yes’ (1) or ‘no’ (0). \\ \midrule
    25 & Has cable services & Can take on values ‘yes’ (1) or ‘no’ (0). \\ \midrule
    26 & Tenure status of the housing unit & Can take on 7 values from ‘own or owner-like possession of the house and lot’ (1), ‘own the house, rent the lot’ (2), ‘own the house, rent-free lot with consent of the owner’ (3), and more. \\ \midrule
    27 & Type of place of residence & Can take on values ‘urban’ (1) or ‘rural’ (2). \\ \midrule
    28 & Time to get to water source (minutes) & Numerical value. \\ \midrule
    29 & Has electricity & Can take on values ‘yes’ (1) or ‘no’ (0). \\ \midrule
    30 & Number of rooms used for sleeping & Numerical value. \\ \midrule
    31 & Has mobile telephone & Can take on values ‘yes’ (1) or ‘no’ (0). \\ \midrule
    32 & Mobile phone used for financial transactions & Can take on values ‘yes’ (1) or ‘no’ (0). \\ \midrule
    33 & Has induction stove & Can take on values ‘yes’ (1) or ‘no’ (0). \\ \midrule
    34 & Has DVD player & Can take on values ‘yes’ (1) or ‘no’ (0). \\ \midrule
    35 & Beneficiary of Pantawid Pamilyang Pilipino Program (4Ps) & Can take on values ‘yes’ (1) or ‘no’ (0). \\ \midrule
    36 & Observed place for handwashing & Can take on 6 different values from ‘observed, fixed facility (sink/tap) in dwelling’ (1), ‘observed, fixed facility (sink/tap) in yard/plot’ (2), ‘observed, mobile object (bucket/jug/kettle)’ (3), and more. \\ 
\end{longtable}
\endgroup

\subsection{Machine Learning Models}
This study employed five boosting algorithms: Adaptive Boosting (AdaBoost), Cat Boosting (CatBoost), Gradient Boosting Machine (GBM), Light Gradient Boosting Machine (LightGBM), and Extreme Gradient Boosting (XGBoost). These models were chosen for their robustness and their ability to handle high-dimensional data, which is crucial for a study that aimed to consider the multidimensional nature of poverty. To handle class imbalance, the Synthetic Minority Over-sampling Technique (SMOTE) was employed on the training data. Hyperparameter tuning was conducted both using manual trial and error and grid search on the validation data. Table 2 showcases the hyperparameters for the project.

\begin{longtable}{lll}
    \caption{Parameters for Different Gradient Boosting Algorithms} \\
    \toprule
    \textbf{Algorithm} & \textbf{Parameter} & \textbf{Value} \\
    \midrule
    \endfirsthead
    \midrule
    \textbf{Algorithm} & \textbf{Parameter} & \textbf{Value} \\
    \midrule
    \endhead
    \bottomrule
    \endfoot
    \bottomrule
    \endlastfoot
    Adaptive Boosting (AdaBoost) & learning\_rate & 0.5 \\
     & n\_estimators & 200 \\
    \midrule
    Cat Boosting (CatBoost) & depth & 4 \\
     & iterations & 300 \\
     & learning\_rate & 0.3 \\
    \midrule
    Gradient Boosting Machine (GBM) & learning\_rate & 0.3 \\
     & max\_depth & 3 \\
     & n\_estimators & 300 \\
    \midrule
    Light Gradient Boosting Machine (LightGBM) & learning\_rate & 0.1 \\
     & n\_estimators & 200 \\
     & num\_leaves & 31 \\
    \midrule
    Extreme Gradient Boosting (XGBoost) & learning\_rate & 0.3 \\
     & max\_depth & 3 \\
     & n\_estimators & 300 \\
\end{longtable}

\subsection{Performance Metrics}
To evaluate the performance of various machine learning algorithms in predicting and classifying household poverty levels, a range of performance metrics was calculated and compared to thoroughly evaluate the models. The performance of each algorithm was assessed based on the average of the metrics. These metrics include:

\textbf{Accuracy Score.} The accuracy score is the ratio of correctly predicted instances to the total instances in the dataset. It is calculated as the sum of true positives and true negatives divided by the total number of predictions:
\begin{equation}
\text{Accuracy} = \frac{TP + TN}{TP + TN + FP + FN}
\end{equation}
where TP represents true positives (i.e., poor households correctly identified as poor), TN represents true negatives (i.e., non-poor households correctly identified as non-poor), FP represents false positives (i.e., non-poor households incorrectly identified as poor), and FN represents false negatives (i.e., poor households incorrectly identified as non-poor).

\textbf{Precision Score.} Precision measures the proportion of correctly predicted positive instances to the total predicted positive instances. It is calculated as:
\begin{equation}
\text{Precision} = \frac{TP}{TP + FP}
\end{equation}

\textbf{Recall.} Recall, also known as sensitivity or the true positive rate, measures the proportion of correctly predicted positive instances to all actual positive instances. It is calculated as:
\begin{equation}
\text{Recall} = \frac{TP}{TP + FN}
\end{equation}

\textbf{F1 Measure.} The F1 score is the harmonic mean of precision and recall and provides a balanced measure that takes both into account. It is calculated as:
\begin{equation}
\text{F1 Score} = 2 \times \frac{\text{Precision} \times \text{Recall}}{\text{Precision} + \text{Recall}}
\end{equation}

\textbf{Area Under the Receiver Operating Characteristic Curve (AUC-ROC).} The ROC, or Receiver Operating Characteristic curve, plots the recall against the FP Rate (FPR) at various threshold settings. The AUC, or Area Under the Curve, quantifies the overall performance of the model by calculating the area under the ROC curve, with values ranging from 0 to 1. A model with an AUC closer to 1 indicates excellent performance and effectively distinguishes between positive and negative classes, while an AUC closer to 0.5 suggests the model performs no better than random guessing.

\textbf{Confusion Matrix.} A confusion matrix is a tabular representation of actual versus predicted classifications, composed of true positives, true negatives, false positives, and false negatives. This matrix provides an overview of model performance through helping identify specific types of classification errors. While the confusion matrix provides detailed data, it can be less informative in high-dimensional problems with many classes and requires careful interpretation to understand the significance of each component. Metrics like accuracy, precision, and recall counterbalance the confusion matrix by summarizing these details into more interpretable forms.

The computational efficacy of various metrics in addition to traditional performance metrics above was also examined. The computational efficacy was obtained by measuring the training time, testing time, and model size. To capture these metrics, the time module was utilized to record the duration of training and testing processes, and the memory\_profiler library to monitor the memory usage and model size. These additional assessments provided a comprehensive understanding of the efficiency and practicality of the models in real-world applications.

\subsection{Tools and Libraries}
The study employed various Python libraries and software for data analysis and machine learning operations. Specifically, the study utilized NumPy (version 1.25.0) and pandas (version 1.5.1) for data manipulation and preprocessing, seaborn (version 0.13.2) and matplotlib (version 3.6.0) for data visualization, and scipy (version 1.10.0) for statistical analysis. For machine learning modeling and evaluation, the study used scikit-learn (version 1.4.2), lightgbm (version 4.2.0), xgboost (version 1.8.0), and catboost (version 1.1.0). Additionally, memory\_profiler (version 0.61.0) was utilized to monitor and optimize memory usage during the implementation. These packages are accessible via the website Python Package Index or through pip.

\section{Results}

The research employed five machine learning models on DHS data from the Philippines: Adaptive Boosting (AdaBoost), Cat Boosting (CatBoost), Gradient Boosting Machine (GBM), Light Gradient Boosting Machine (LightGBM), and Extreme Gradient Boosting (XGBoost). Wealth classification was approached as a five-class problem (Richest, Richer, Middle, Poorer, Poorest). Each model was trained on 80\% of the cleaned dataset and evaluated on the remaining 20\%. Hyperparameters were optimized using a validation set, which comprised 10\% of the training data, with performance assessed through metrics such as accuracy, precision, recall, and F1-score.

As shown in Table 3, the performance metrics for the five models were as follows: CatBoost achieved the highest accuracy at 90.93\%, followed by XGBoost at 89.41\%, GBM at 89.05\%, and LightGBM at 88.52\%. AdaBoost had the lowest accuracy at 80.39\%. In terms of precision, CatBoost again ranked highest with 90.92\%, followed by XGBoost at 89.39\%, GBM at 89.04\%, and LightGBM at 88.51\%, with AdaBoost recording a precision of 83.55\%. For recall, CatBoost led with 90.93\%, while XGBoost had 89.41\%, GBM showed 89.05\%, and LightGBM was close behind with 88.52\%. AdaBoost had the lowest recall at 80.39\%. Regarding the F1 score, CatBoost achieved the highest at 90.92\%, followed by XGBoost with 89.40\%, GBM at 89.04\%, and LightGBM at 88.50\%, with AdaBoost recording the lowest F1 score at 80.15\%. The rankings of the models with respect to all metrics were consistent: CatBoost was first, followed by XGBoost, GBM, LightGBM, and AdaBoost in that order. However, the metric values for the top four models are remarkably similar, with only AdaBoost exhibiting significantly lower performance metrics.

\begin{longtable}{lllll}
    \caption{Performance Evaluation Metrics of the Boosting Models} \\
    \toprule
    \textbf{Model} & \textbf{Accuracy} & \textbf{Precision} & \textbf{Recall} & \textbf{F1 Score} \\
    \midrule
    \endfirsthead
    \midrule
    \textbf{Model} & \textbf{Accuracy} & \textbf{Precision} & \textbf{Recall} & \textbf{F1 Score} \\
    \midrule
    \endhead
    \bottomrule
    \endfoot
    \bottomrule
    \endlastfoot
    AdaBoost & 0.803917 & 0.835551 & 0.803917 & 0.801523 \\
    CatBoost & 0.909333 & 0.909193 & 0.909333 & 0.909191 \\
    GBM & 0.890474 & 0.890362 & 0.890474 & 0.890353 \\
    LightGBM & 0.885155 & 0.885137 & 0.885155 & 0.884996 \\
    XGBoost & 0.894101 & 0.893919 & 0.894101 & 0.893981 \\
\end{longtable}

Meanwhile, Table 4 provides the AUC-ROC scores to assess the performance of different models in predicting household poverty levels. For the "Poorest" class, CatBoost, GBM, LightGBM, and XGBoost achieved scores around 0.98 to 0.99, while AdaBoost scored significantly lower at 0.90. This is reflected in the confusion matrix for AdaBoost (Figure 1), which shows a higher number of misclassifications, particularly with true "Poorest" instances being predicted as "Poorer" and some as "Richer." On the other hand, for the "Poorer" class, CatBoost, GBM, LightGBM, and XGBoost all achieved high scores of 0.99, while AdaBoost scored lower at 0.73. AdaBoost's confusion matrix reveals significant misclassifications for the "Poorer" class, with many instances being misclassified as "Poorest" or "Middle." In both classes, CatBoost, GBM, LightGBM, and XGBoost show fewer misclassifications (FigureS 2.2-2.5), consistent with their higher AUC-ROC scores. For the "Middle" class, all models demonstrated perfect performance with scores of 1.00, which is confirmed by their confusion matrices showing almost no misclassifications for this class.

In the "Richer" class, CatBoost, GBM, LightGBM, and XGBoost again achieved perfect scores of 1.00, while AdaBoost scored lower at 0.79. AdaBoost's confusion matrix shows misclassifications for the "Richer" class, with some instances being misclassified as "Poorer" or "Richest." Conversely, CatBoost, GBM, LightGBM, and XGBoost showed strong performance with minimal misclassifications. Lastly, for the "Richest" class, CatBoost, GBM, LightGBM, and XGBoost achieved near-perfect scores of 1.00, and AdaBoost scored slightly lower at 0.99. The confusion matrix for AdaBoost shows a small number of misclassifications for this class, with some instances predicted as "Richer."

\begin{longtable}{lccccc}
    \caption{AUC-ROC Scores for Each Class Across the Boosting Models} 
    \label{tbl:auc_roc_scores} \\
    \toprule
    \textbf{Class} & \textbf{AdaBoost} & \textbf{CatBoost} & \textbf{GBM} & \textbf{LightGBM} & \textbf{XGBoost} \\
    \midrule
    \endfirsthead
    \midrule
    \textbf{Class} & \textbf{AdaBoost} & \textbf{CatBoost} & \textbf{GBM} & \textbf{LightGBM} & \textbf{XGBoost} \\
    \midrule
    \endhead
    \bottomrule
    \endfoot
    \bottomrule
    \endlastfoot
    Poorest & 0.90 & 0.99 & 0.98 & 0.98 & 0.98 \\
    Poorer & 0.73 & 0.99 & 0.99 & 0.99 & 0.99 \\
    Middle & 0.99 & 1.00 & 1.00 & 1.00 & 1.00 \\
    Richer & 0.79 & 0.99 & 0.99 & 0.99 & 0.99 \\
    Richest & 1.00 & 1.00 & 1.00 & 1.00 & 1.00 \\
\end{longtable}

\begin{figure}[ht]
    \centering
    \begin{minipage}{0.45\textwidth}
        \centering
        \includegraphics[width=\textwidth]{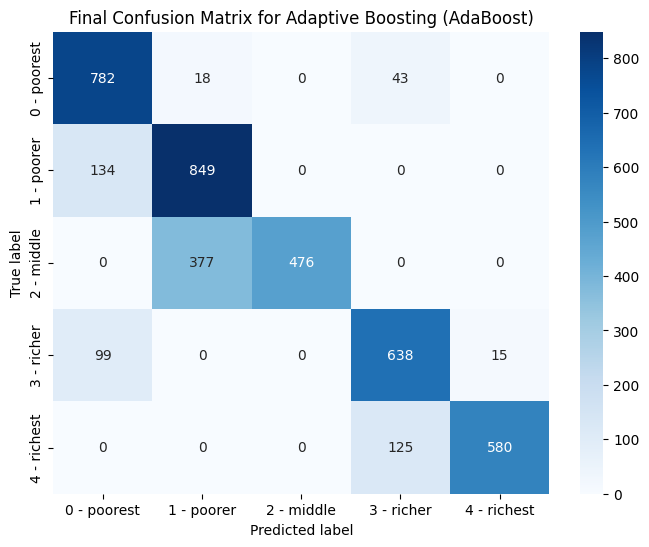}
    \end{minipage}
    \hfill
    \begin{minipage}{0.45\textwidth}
        \centering
        \includegraphics[width=\textwidth]{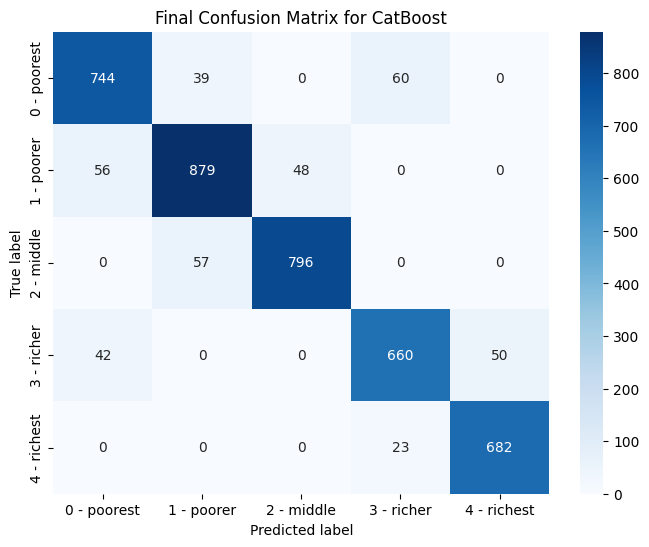}
    \end{minipage}
    
    \vspace{1em} 
    
    \begin{minipage}{0.45\textwidth}
        \centering
        \includegraphics[width=\textwidth]{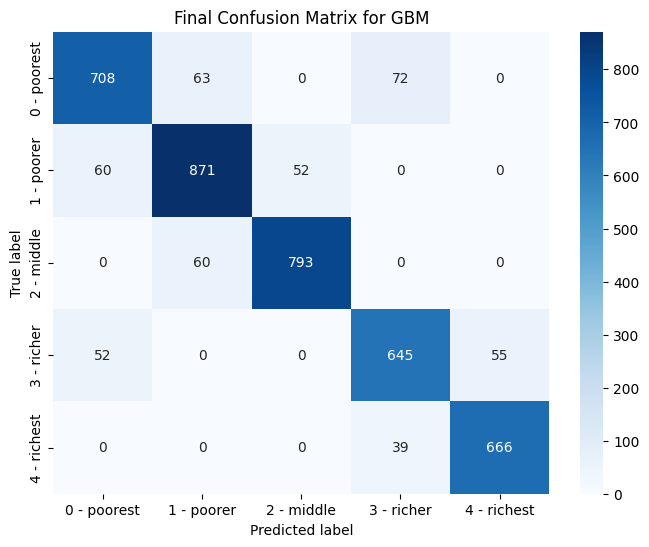}
    \end{minipage}
    \hfill
    \begin{minipage}{0.45\textwidth}
        \centering
        \includegraphics[width=\textwidth]{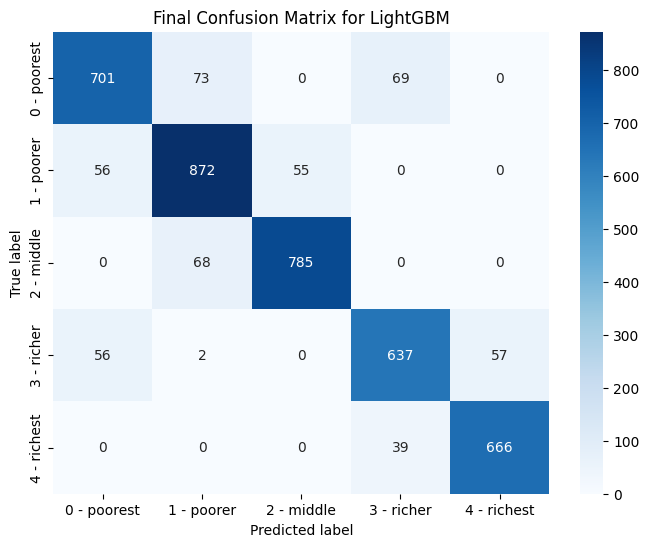}
    \end{minipage}
    
    \vspace{1em} 
    
    \begin{minipage}{0.45\textwidth}
        \centering
        \includegraphics[width=\textwidth]{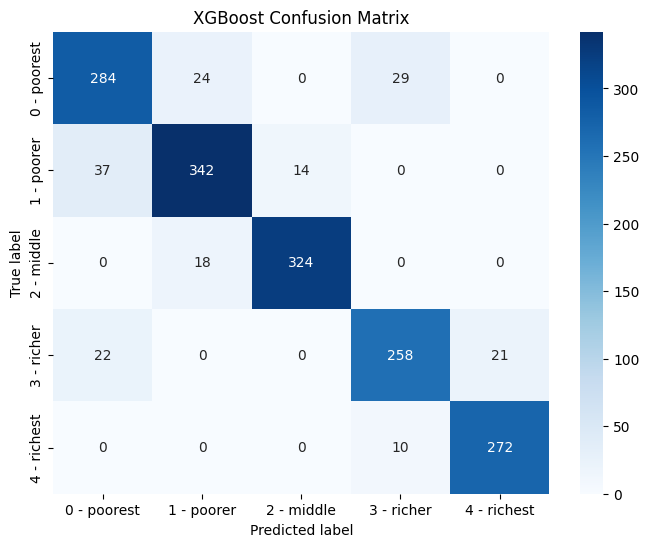}
    \end{minipage}
    \caption{Figures 2.1-2.5 (From Left to Right): Confusion Matrices for AdaBoost (Fig 2.1), CatBoost (Fig 2.2), GBM (Fig 2.3), LightGBM (Fig 2.4), and XGBoost (Fig 2.5).}
    \label{fig:confusion_matrices}
\end{figure}

Additionally, the study examined the models’ computational efficacy to assess the practical applicability of the models. Recent research often neglects these metrics and focuses instead on previous benchmarks. Table 5 provides the comparison of various models in terms of training time, testing time, and model size. AdaBoost stands out with the shortest training time at approximately 4.48 seconds, making it the quickest model to train. However, it takes the longest time for testing at 0.23 seconds. In contrast, CatBoost, while having the longest training time of 69.29 seconds and the largest model size at 30.50 MB, demonstrates exceptional efficiency during testing, taking only 0.01 seconds. GBM (Gradient Boosting Machine) shows moderate performance, with a training time of 16.81 seconds, a testing time of 0.02 seconds, and a model size of 15.80 MB. LightGBM and XGBoost exhibit a good balance, featuring relatively quick training times of 2.17 and 2.58 seconds, respectively, along with small model sizes of 2.50 MB and 3.10 MB. LightGBM takes slightly longer during testing at 0.07 seconds, compared to XGBoost's 0.03 seconds.

\begin{longtable}{lccccc}
    \caption{Computational Efficacy Metrics Across the Boosting Models}
    \label{tbl:model_performance_metrics} \\
    \toprule
    \textbf{Metric} & \textbf{AdaBoost} & \textbf{CatBoost} & \textbf{GBM} & \textbf{LightGBM} & \textbf{XGBoost} \\
    \midrule
    \endfirsthead
    \midrule
    \textbf{Metric} & \textbf{AdaBoost} & \textbf{CatBoost} & \textbf{GBM} & \textbf{LightGBM} & \textbf{XGBoost} \\
    \midrule
    \endhead
    \bottomrule
    \endfoot
    \bottomrule
    \endlastfoot
    Training Time (seconds) & 4.48 & 69.29 & 16.81 & 2.17 & 2.58 \\
    Testing Time (seconds) & 0.23 & 0.01 & 0.02 & 0.07 & 0.03 \\
    Model Size (MB) & 1.20 & 30.50 & 15.80 & 2.50 & 3.10 \\
\end{longtable}

\section{Discussion}

The evaluation of five machine learning models—AdaBoost, CatBoost, GBM, LightGBM, and XGBM—on DHS data from the Philippines revealed that CatBoost consistently achieved the highest performance metrics, including accuracy, precision, recall, and F1-score. XGBoost followed closely, with GBM and LightGBM also demonstrating strong performance. AdaBoost lagged behind with the lowest performance across all metrics. Moreover, AUC-ROC curves further validated the models' discriminative capabilities in predicting household poverty levels. CatBoost, GBM, LightGBM, and XGBoost achieved near-perfect AUC values across most classes, particularly in distinguishing the "Poorest," "Middle," "Richer," and "Richest" classes. AdaBoost showed significantly lower AUC scores, especially for the "Poorest" and "Poorer" classes, which was reflected in its higher misclassification rates. In terms of computational efficiency, AdaBoost had the shortest training time but the longest testing time. CatBoost required the longest training time and the largest model size but demonstrated exceptional testing efficiency. GBM, LightGBM, and XGBoost balanced well between training and testing times, with LightGBM and XGBoost also showing smaller model sizes.

Overall, CatBoost emerged as the top performer across all metrics, followed closely by XGBoost, GBM, and LightGBM. AdaBoost, while efficient in training time, showed lower performance in accuracy, precision, recall, and F1-score, as well as higher misclassification rates. LightGBM and XGBoost demonstrated a good balance of high performance and computational efficiency, thus are strong candidates for practical applications.

This study also highlighted the most impactful features in predicting poverty through the feature selection method, as outlined in Table 1, indirectly suggesting potential areas for policy focus. Future research could model how changes in these features affect predicted poverty classes explicitly. These findings have global implications for poverty alleviation efforts. Policymakers can gain more accurate insights into poverty dynamics and develop targeted interventions addressing the multifaceted nature of poverty by utilizing machine learning techniques. However, the limitations in this study, such as the reliance on DHS data and the need for further validation using alternative datasets or methodologies must be acknowledged. Therefore, incorporating more complex types of information for analysis and poverty prediction is necessary. Combining GPS data with survey data, for instance, could significantly enhance the accuracy of poverty level classification in the Philippines. Utilizing more sophisticated information, such as GPS data, night light data, and other advanced metrics, could improve the precision of poverty predictions.

\section{Conclusion}
This study demonstrated the effectiveness of machine learning boosting algorithms, particularly CatBoost, in predicting household poverty levels in the Philippines. CatBoost emerged as the top performer by offering high accuracy and computational efficiency. However, AdaBoost lagged behind in performance metrics. Feature selection highlighted areas for potential policy intervention. Overall, this research contributes development to poverty alleviation efforts through the utilization of advanced technology.

\section{Bibliography}

\bibliographystyle{unsrtnat}
\bibliography{ref}

\end{document}